\documentclass[11pt]{article}
\usepackage[utf8]{inputenc}
\usepackage{hyperref}
\usepackage{xcolor}
\usepackage{authblk}

\usepackage{fullpage}

\usepackage{graphicx,lscape,rotating}

\usepackage{amssymb}
\usepackage{amsmath}
\usepackage{multirow}
\newcommand\EatDot[1]{}

\newcommand\snowmass{\begin{center}\rule[-0.2in]{\hsize}{0.01in}\\\rule{\hsize}{0.01in}\\
\vskip 0.1in Submitted to the  Proceedings of the US Community Study\\ 
on the Future of Particle Physics (Snowmass 2021)\\ 
\rule{\hsize}{0.01in}\\\rule[+0.2in]{\hsize}{0.01in} \end{center}}

\newcommand{\catania}{INFN Sezione di Catania, I-95123 Catania, Italy}

\newcommand{\ceem}{Center for  Exploration  of  Energy  and  Matter,
Indiana  University,
Bloomington,  IN  47403,  USA}

\newcommand{\icn}{Instituto de Ciencias Nucleares, 
Universidad Nacional Aut\'onoma de M\'exico, Ciudad de M\'exico 04510, Mexico}
\newcommand{\icsup}{Pedagogical University of Krak\'ow, 30-084 Krak\'ow, Poland}
\newcommand{\indiana}{Department of Physics,
Indiana  University, Bloomington,  IN  47405,  USA}
\newcommand{\jlabth}{Theory Center, Thomas  Jefferson  National  Accelerator  Facility, Newport  News,  VA  23606,  USA}

\newcommand{\ucm}{Departamento de F\'isica Te\'orica, Universidad Complutense de Madrid and IPARCOS, 28040 Madrid, Spain}
\newcommand{\uned}{Departamento de F\'isica Interdisciplinar, Universidad Nacional de Educaci\'on a Distancia (UNED), Madrid E-28040, Spain}

\newcommand{\messina}{Dipartimento di Scienze Matematiche e Informatiche, Scienze Fisiche e Scienze della Terra, 
Universit\`a degli Studi di Messina, I-98122 Messina, Italy}

\newcommand{\scnuIQM}{Guangdong Provincial Key Laboratory of Nuclear Science, Institute of Quantum Matter, South China Normal University, Guangzhou 510006, China}
\newcommand{\scnuJLQM}{Guangdong-Hong Kong Joint Laboratory of Quantum Matter, Southern Nuclear Science Computing Center, South China Normal University, Guangzhou 510006, China}
\newcommand{\ub}{Departament de F\'isica Qu\`antica i Astrof\'isica and Institut de Ci\`encies del Cosmos, Universitat de Barcelona, E-08028, Spain}
\newcommand{\ific}{Instituto de F\'isica Corpuscular (IFIC), Centro Mixto CSIC-Universidad de Valencia, E-46071 Valencia, Spain}
\newcommand{\fsu}{Florida State University, Tallahassee, Florida 32306, USA}
\newcommand{\wm}{William \& Mary, Williamsburg, Virginia 23185, USA}

\newcommand{\tubingen}{Institute for Theoretical Physics, T\"ubingen University, 72076 T\"ubingen, Germany}

\title{ Hadron Spectroscopy in Photoproduction }

\author[1]{Miguel~Albaladejo}
\affil[1]{\ific}

\author[2]{{\L}ukasz~Bibrzycki}
\affil[2]{\icsup}

\author[3]{Sean Dobbs}
\affil[3]{\fsu}

\author[4,5]{C\'esar~Fern\'andez-Ram\'irez}
\affil[4]{\uned}
\affil[5]{\icn}

\author[6]{Astrid~N.~Hiller~Blin}
\affil[6]{\tubingen}

\author[7,8]{Vincent~Mathieu}
\affil[7]{\ub}
\affil[8]{\ucm}

\author[9,10]{Alessandro~Pilloni}
\affil[9]{\messina}
\affil[10]{\catania}

\author[11]{Justin Stevens}
\affil[11]{\wm}

\author[12,13,14]{Adam~P.~Szczepaniak}
\affil[12]{\jlabth}
\affil[13]{\ceem}
\affil[14]{\indiana}

\author[13,14,15,16]{Daniel~Winney}
\affil[15]{\scnuJLQM}
\affil[16]{\scnuIQM}

\date{\today}

\begin{document}

\snowmass

{\let\newpage\relax\maketitle}. 


\section{Introduction \& Summary}

Recent decades have seen a resurge of interest in hadron spectroscopy, driven by new, high-luminosity experiments which among other achievements have led to the discovery of dozens of new heavy-quark hadrons which do not fit well into the spectrum of states from the quark model, the so-called ``$XYZP$'' states.  Although the existence of these states strongly suggest hadronic degrees of freedom beyond the naive quark model, their interpretation has been limited by our limited knowledge of the properties of these states, their masses, widths, and $J^{PC}$ quantum numbers.  The existence of many of these states is also an open question, particularly for those which have only been observed by one experiment.   Photoproduction has emerged as an attractive process to study hadron spectroscopy, due to the range of states accessible in new and planned experimental facilities, complimentary kinematics to other experiments where rescattering effects near thresholds are reduced, and advances in our theoretical understanding of these reactions.

Historically, most photoproduction experiments have been at photon beam energies of a few GeV, particularly with a focus on light quark baryon spectroscopy.
Previous measurements in $ep$ collisions at HERA, however, have demonstrated the ability to study heavy quarkonia through photoproduction, particularly the well known vector $c\bar{c}$ and $b\bar{b}$ states~\cite{Adloff:2002re,Alexa:2013xxa,Adloff:2000vm,Chekanov:2009zz}.  The COMPASS collaboration has studied muonproduction of the $J/\psi \pi^+\pi^- p$ final state finding an indication of a new state $\widetilde{X}(3872)$~\cite{Aghasyan:2017utv} and also set limits on $Z_c$ photoproduction in the $J/\psi \pi^+ n$ final state~\cite{Adolph:2014hba}.  

This white paper reviews the prospects for hadron spectroscopy from three major existing and proposed facilities:
\begin{itemize}
    \item \textit{GlueX} started taking data in 2017 with a linearly polarized photon beam of up to 12~GeV in energy.  The focus of the experiment is on the search for light quark hybrid mesons, and an active program of amplitude analysis is ongoing with preliminary results on the channels $\eta\pi$ and $\omega\pi$ available. A luminosity of $840$~pb$^{-1}$ has been collected for beam energies $>6$~GeV, with a factor of more than 2 increase expected by 2025. The large data sets enable the development of detailed reaction models and searches for strange quark analogues of the $XYZP$ states.  GlueX also can study the photoproduction of charmonia up to the $\psi(2S)$, particularly final states including a $J/\psi\to e^+e^-$ decay, including the $s$-channel production of $P_c^+$ states. 
    \item The \textit{Electron-Ion Collider} is a high-luminosity polarized $ep$ and $eA$ collider with variable center-of-mass energies from 20-140 GeV, which is planned to start taking data after 2030.  Integrated luminosities are expected to be on the order of 1-10~fb$^{-1}$, so there are many opportunities for studying $XYZ$ states.  Simulation studies have been performed for $\gamma p \to Z_c(3900)^+n$ and seem very promising.
    \item A proposed \textit{JLab24} upgrade would increase the maximum energy of the CEBAF accelerator at Jefferson Lab to 24~GeV, which would provide efficient production of $X$ and $Z$ states which are expected to be enhanced near there production threshold.  The fixed-target luminosities are several orders of magnitude larger than those accessible at the EIC, providing access to more rare processes.
\end{itemize}
Additional information is expected to come from quasi-real photoproduction measurements at CLAS12 at Jefferson Lab, and other shorter, dedicated experiments such as the $J/\psi$-007~experiment.


\section{Theory}
\label{sec:jpac}

Real and virtual photons are efficient probes of the internal structure of hadrons. Their collisions with hadron targets result in a copious production of meson and baryon resonances. 
Resonance properties directly reflect into the dependence of their photoproduction observables on the exchange momentum and photon virtuality. Measuring these with high precision in a large kinematic range will thus provide valuable insights on the nature of the excited states of QCD. 

This is particularly relevant for the study of the  new resonance candidates, commonly  referred to as the XYZ, that populated the heavy quarkonium spectrum since 2003~\cite{Esposito:2016noz,Guo:2017jvc,Olsen:2017bmm,Brambilla:2019esw}. 
Their properties do not fit the expectations for heavy $Q \bar Q$ bound states as predicted by the conventional phenomenology. An exotic composition is most likely required. Having a comprehensive description of these states will improve our understanding of the nonperturbative features of Quantum Chromodynamics. 
None of these states have been uncontroversially seen in electro- or photoproduction yet, which would provide complementary information, that can further shed light on their nature. 
In particular, electro- and photo-production at high energies is not affected by 3-body dynamics which complicates the determination of the resonant nature of several XYZ~\cite{Szczepaniak:2015eza,Albaladejo:2015lob,Guo:2016bkl,Pilloni:2016obd,Nakamura:2019btl,Guo:2019twa}. 
The theoretical framework to calculate electromagnetic transitions of conventional  quarkonia is rather robust, which makes the predictions for photoproduction observables particularly reliable, and be the benchmark for the XYZ as well. 

The various interpretations of the XYZ states differ for their spatial extension, a hadron molecule having a bigger radius than a quark state bound by color forces. The size of a resonance can be probed in electroproduction by sitting at the resonance pole and extracting the dependence on the photon virtuality. The latter can be compared with theoretical models to provide insight of nature of the state, as for example done for the Roper~\cite{Segovia:2015hra}, or to assess the existence of hybrid baryons~\cite{Segovia:2019jdk}. 


The identification of an exotic state requires not only precise data but also an equally refined amplitude parametrization. Most of the extractions of XYZ resonance parameter rely on simplistic models that may lead to misinterpretation of their nature. For example, that was the case for the light hybrid sector, where the $\pi_1(1400)$ and $\pi_1(1600)$ were considered different states decaying independently to different final states. A recent coupled-channel analysis of $\eta^{(\prime)}\pi$ based on the general principles of $S$-matrix concluded to the existence of a single state~\cite{JPAC:2018zyd}. Similar cases of apparent duplication of levels might affect the XYZ sector as well, which has immediate consequences for the identification of the multiplets and eventually the understanding of their underlying dynamics.

Dualities play an important role in hadron spectroscopy. Conventional resonances are known to be dual to Regge exchanges in the cross channels, which allows us to extract radiative couplings from measured braching ratios, and use them to predict photoproduction cross sections.  These relations can be formalized using the $S$-matrix principles, as analyticity and dispersion relations. The latter put constraints between the low-energy region of a reaction, which is populated by resonances, and the high-energy region, which is well described by Regge theory. These relations are called Finite Energy Sum Rules (FESR), and are useful to put further theory input to the parametrizations of the low-energy region, which reflects into a more robust and precise determination of the resonance properties. Photoproduction experiments can usually access both the low- and high-energy region at once, facilitating the implementation of FESR in data fits. While single meson photoproduction has been thoroughly explored already~\cite{JPAC:2016lnm,Mathieu:2018mjw}, the application to two- and multi-hadrons is still preliminary~\cite{Shi:2014nea, Bibrzycki:2021rwh} and will require a substantial effort in the coming years.

In addition to reconstructing photoproduction exclusively, the study of semi-inclusive photo- and electroproduction can also be studied with similar tools, depending on the kinematic region of interests. While the peripheral production can be access using duality, the central production at high photon virtuality enters the realm of perturbative QCD. Measurements of prompt $X(3872)$ productions at hadron colliders are already available, and comparing them with electroproduction cross section will provide 
an independent determination of the long-distance matrix elements, and therefore a crosscheck of the underlying model assumptions~\cite{Artoisenet:2009wk, Artoisenet:2010va}.

\begin{figure*}
        \centering
        \raisebox{.8cm}{\includegraphics[width=0.30\textwidth]{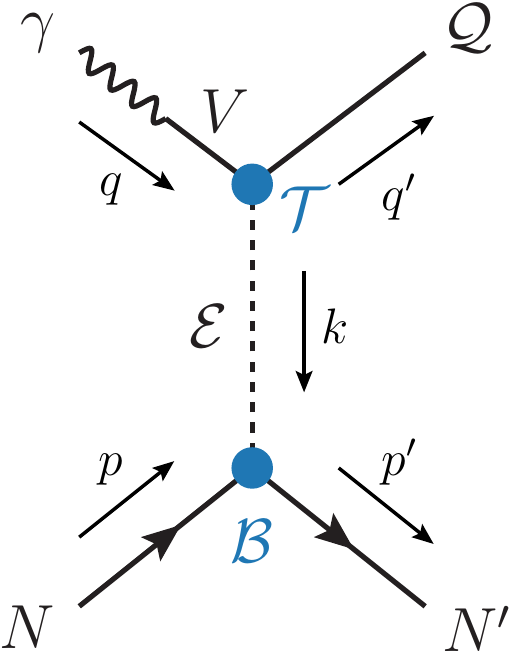} }\hspace{1cm}
        \includegraphics[width=0.5\textwidth]{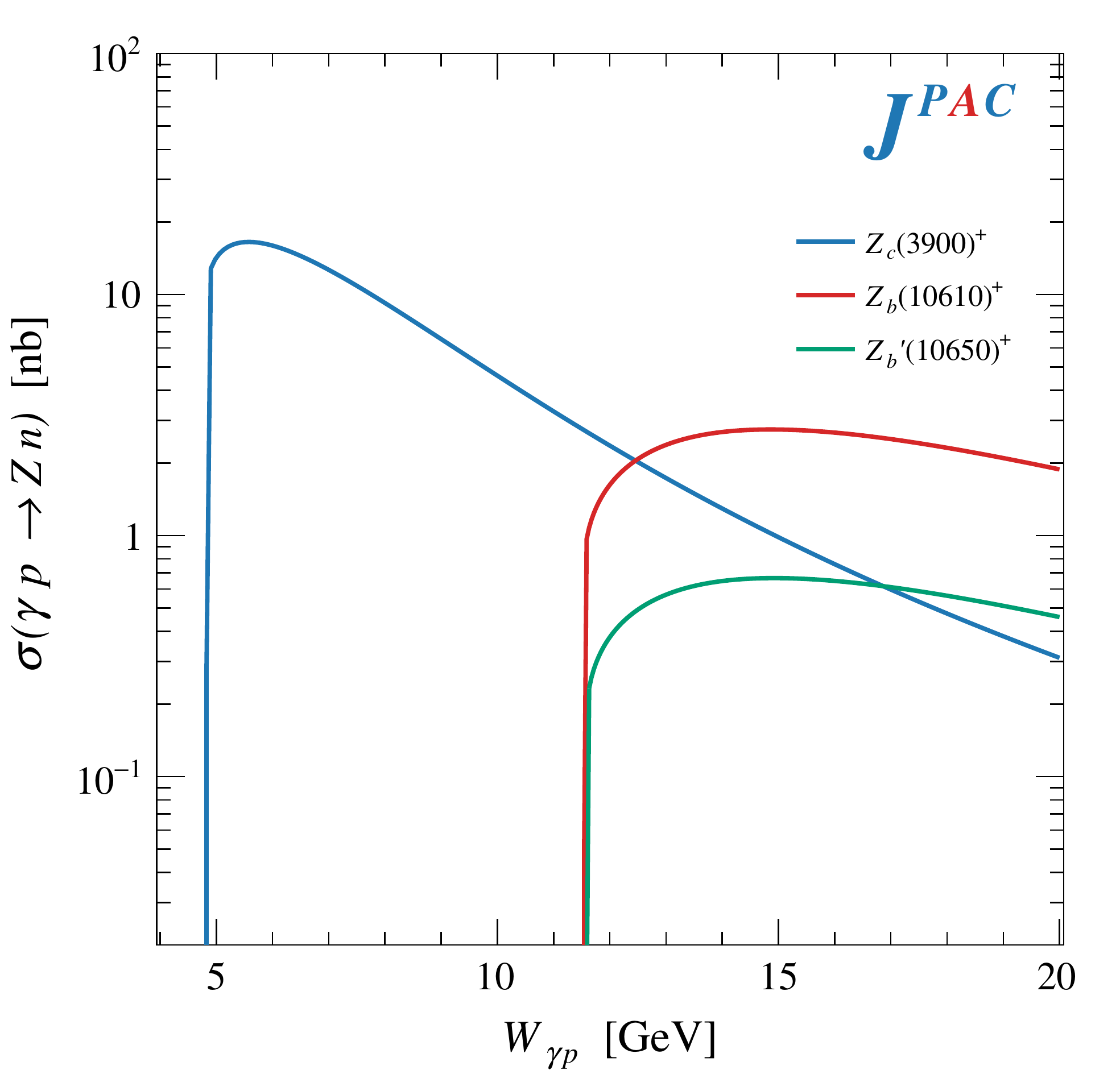}
        
        \caption{{\it Left}, Photoproduction of a quarkonium-like meson, $\mathcal{Q}$ via Reggeon exchange in the $t$-channel ;
        {\it right}, Integrated cross sections for three exotic $Z$ states considered.}
        \label{fig:Zplots}
\end{figure*}


\section{GlueX: Light quark exotics and charm at threshold}
\label{sec:gluex}

The GlueX experiment is a photoproduction experiment located in Hall D of Jefferson Lab in Newport News, VA, which features an intense, multi-GeV linearly polarized photon beam along with a solenoidal spectrometer with nearly full acceptance for charged and neutral particles~\cite{GlueX:2020idb}.  The primary goal of GlueX is the establishment and systematic study of the spectrum of light quark hybrid mesons, that is, mesons in which the excited gluonic field of the meson contributes directly to its properties.  In addition to the study of hybrid mesons, the large acceptance and open trigger of the experiment make a large range of topics in hadronic physics accessible at GlueX.  The maximum accessible photon beam energy of 12 GeV also allows the study of the photoproduction of the bound charmonium states near threshold and the search for the $s$-channel production of states with a possible 5-quark structure like the LHCb $P_c$ states.  Beyond the hybrid meson search, we can identify several major contributions to the broader hadron spectroscopy effort: (1) analyzing the high-statistical precision GlueX data requires the development of sophisticated amplitude models of the photoproduction reactions, and such models of 2- and 3-meson production can be applied to other current and upcoming experiments; (2) the large data set available for the study of strangeonia and strange quark hadrons also allows for the search of strange quark analogues to the $XYZP$ states; (3) the study of charmonium production near threshold allows validation of the photoproduction models described in the previous section and allows detailed searches for the $P_c$ states.

First, we briefly review the recent history of the GlueX experiment and the approved future configurations, which cover the expected experimental program to near the end of the current decade.  The future prospects for GlueX have been discussed in detail in several recent white papers~\cite{gluex_future,Arrington:2021alx}.
The main approved GlueX spectroscopy program is separated into two phases.  The initial Phase-I ran from 2017 to 2018, while Phase-II which features higher photon beam intensity and upgraded $K/\pi$ separation through the addition of a DIRC detector started running in 2020 and is expected to be complete in 2025.  An upgrade to the inner portion of the forward calorimeter from lead glass to lead tungstate crystals, yielding better energy and position resolution, is expected to be performed in 2023.  Several shorter measurements focusing on other hadronic physics topics, such as $\eta$ Primakoff production~\cite{Somov:2020vcu}, pion polarizabilities~\cite{Lawrence:2013asa}, and physics with nuclear targets~\cite{GlueX:2020dvv}, have been scheduled in between the spectroscopy runs. Additional running with an intense ($\approx10^4/s$) $K_L$ beam~\cite{KLF:2020gai}. and a photon beam with a polarized target~\cite{Dalton:2020wdv} have also been approved.  Depending on the outcome of the analysis of the GlueX-I and -II data, additional spectroscopy running under other condition, such as with a different energy of peak linear polarization or with a deuterium or polarized target could also be beneficial.

Mesons with gluonic excitations, known as ``hybrid mesons'', are expected to exist based on model and lattice QCD calculations, and there has been a long history of experimental searches for these states, particularly those with ``exotic'' $J^{PC}$ quantum numbers that can not be formed by a simple quark-antiquark pair~\cite{Meyer:2015eta}.  The most recent progress has been the determination of the resonance parameters of the lightest hybrid meson, the $\pi_1$, from a coupled channel analysis of the $\eta\pi$ and $\eta'\pi$ final states from pion production at COMPASS~\cite{JPAC:2018zyd}, and the first evidence for the isoscalar partner(s) of the $\pi_1$, the $\eta_1$ ($\eta_1'$), $J/\psi\to\gamma\eta\eta'$ at BES-III~\cite{BESIII:2022riz}.   Early GlueX analyses have focused on measurements of cross sections and polarization observables in order to understand the mechanisms of meson photoproduction at these energies, and to build the understanding of the detector response and the theoretical and technical frameworks needed for the amplitude analyses required to extract the signals for hybrid mesons, which are expected to be small~\cite{Meyer:2015eta}.  

\begin{figure*}
        \centering
        \includegraphics[width=0.45\textwidth]{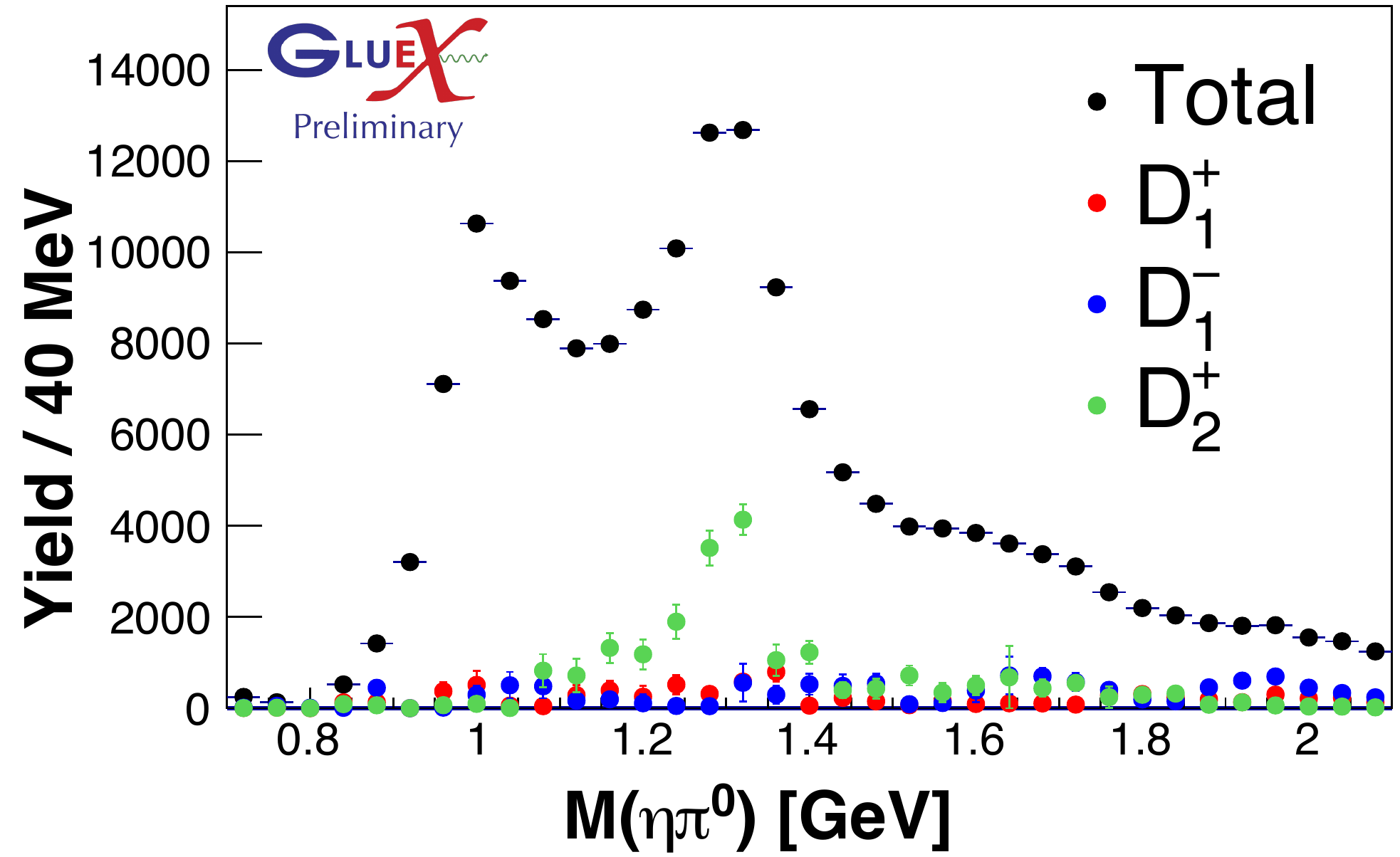}
        \includegraphics[width=0.45\textwidth]{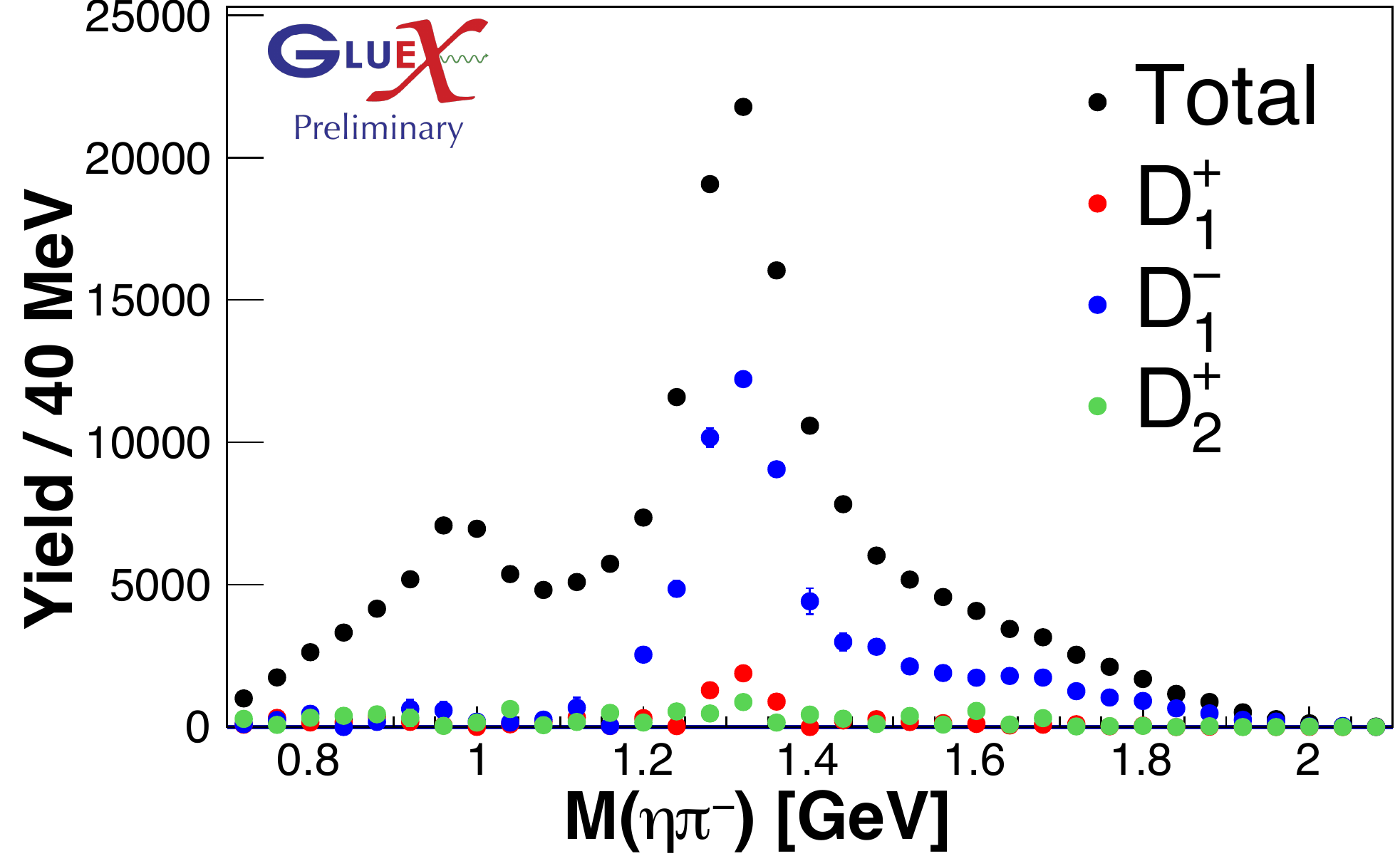}
        \caption{Preliminary GlueX results for the $\eta\pi$ mass spectra and amplitude analysis for the reactions $\gamma p \to \pi^0\eta p$ (left) and $\gamma p \to \eta\pi^- \Delta^{++}$ (right) for $0.1 < -t < 0.3$~GeV$^2$ and $8.2 < E_\gamma < 8.8$~GeV~\cite{Arrington:2021alx}. The black points give the total measured yield and the colored points give the dominant tensor amplitudes from the amplitude analysis, labeled as $L_m^\epsilon$~\cite{Mathieu:2020zpm}.  The production of $a_2(1320)$ is clearly seen in both channels.}
        \label{fig:gluexetapi}
\end{figure*}

Although the $\pi_1$ is expected to decay dominantly to $b_1\pi$~\cite{Woss:2020ayi}, the resulting 5-pion final state is extremely challenging to analyze from both an experimental and theoretical perspective.  We have therefore started to focus on the analysis of final states such as $\eta\pi$ and $\eta'\pi$, which have a much smaller coupling to the $\pi_1$, but are much cleaner to identify, and the amplitude models, while still complex, are much more tractable.  Another early focus is on the $\omega\pi$ final state, which allows the study of the decay $b_1\to\omega\pi$ as a stepping stone to $b_1\pi$ and a search for excited vector states.  These final states have healthy photoproduction cross sections, and allow for complementary investigations, for example studying different decay modes (e.g. $\eta\to\gamma\gamma$ and $\pi^+\pi^-\pi^0$) or differently charged final states states (e.g. $\gamma p \to \omega \pi^0 p$ and $\gamma p \to \omega \pi^- \Delta^{++}$), which extend our understanding of these reactions.  Preliminary results from these analyses have been shown at conferences, for example as shown for the $\eta\pi$ channel in Fig.~\ref{fig:gluexetapi}, and we expect results on these reactions to be published in the near future. 

With these amplitude analysis frameworks established, we expect to be able to apply them directly to the analysis of other two-pseudoscalar final states ($\pi^+\pi^-,\pi^0\pi^0,K^+K^-,K_SK_S,\eta\eta,\eta\eta'$) and other vector-pseudoscalar final states ($\omega\eta,\phi\pi,\phi\omega,\phi\eta,KK^*,KK_1$).  Many of these final states are under active investigation, and promise to provide a wealth of information on normal and exotic mesons.  The comparison of the couplings of the states to strange and non-strange quark final states also promises to given information on their strange-quark content, though the full analysis will surely require the collection of the full GlueX-II data.   Progress is also being made on the theoretical frameworks to analyze more complicated important reactions, such as tensor+pseudoscalar and 3- and 4-body final states.  We expect these to come later in the decade, when the full GlueX-II data is in hand.  Finally, many of these same final states can be used to look for the strange-quark analogues of the $XYZP$ states, such as the possible $Z_s\to\phi\pi$ and $P_s\to\phi p$.   One example is the $\phi\pi^+\pi^-$ final state currently under analysis, which is analogous to the $J/\psi \pi^+\pi^-$ in which several $X$ and $Y$ states have been seen, and can be used to search for the potentially-exotic $\phi(2170)$~\cite{gluex_phipipi_nacer}.  Although this reaction show many interesting structures, to fully understand it we will need a 3-body amplitude model and the full GlueX data.

\begin{figure*}
        \centering
        \includegraphics[width=0.45\textwidth]{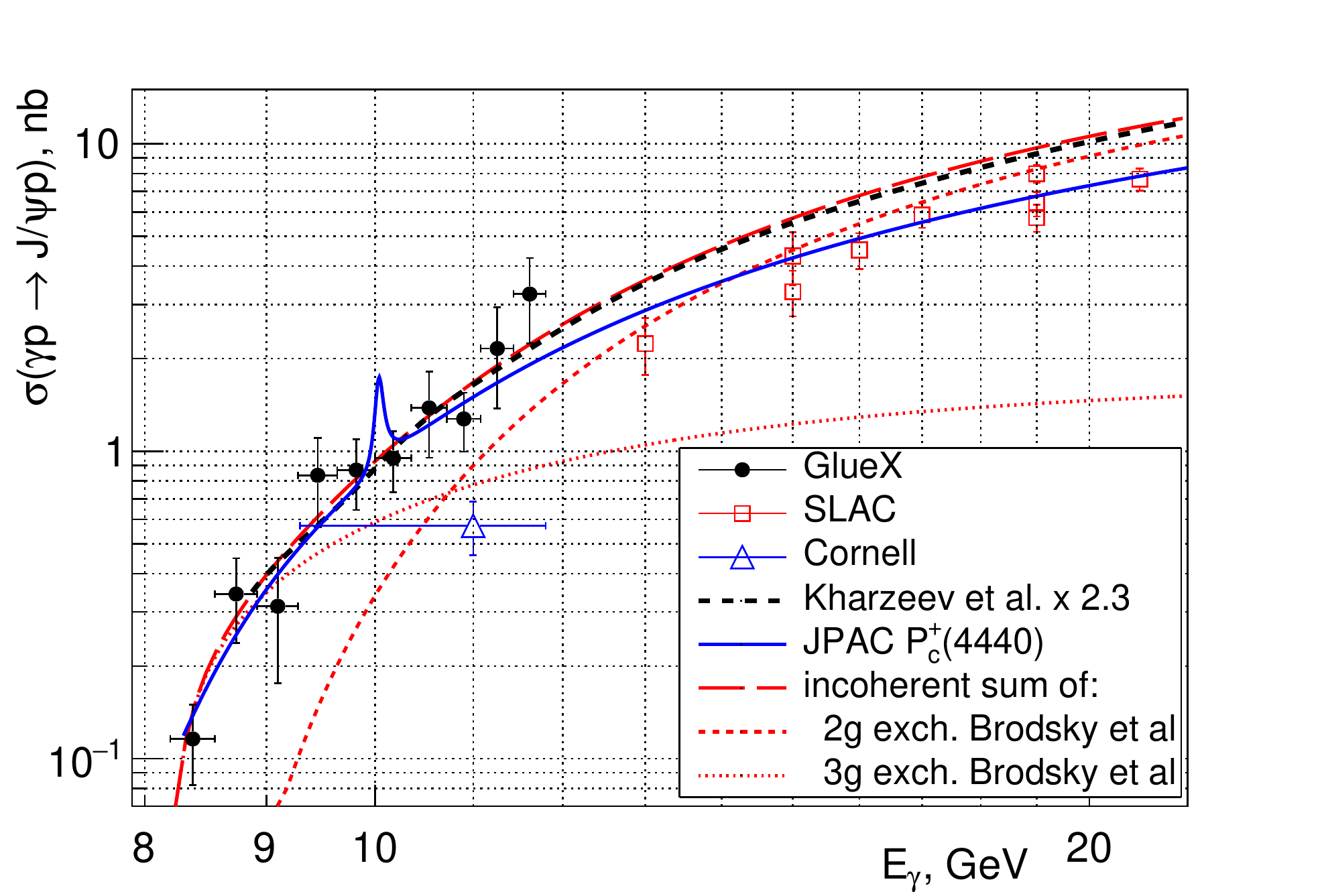}
        \includegraphics[width=0.45\textwidth]{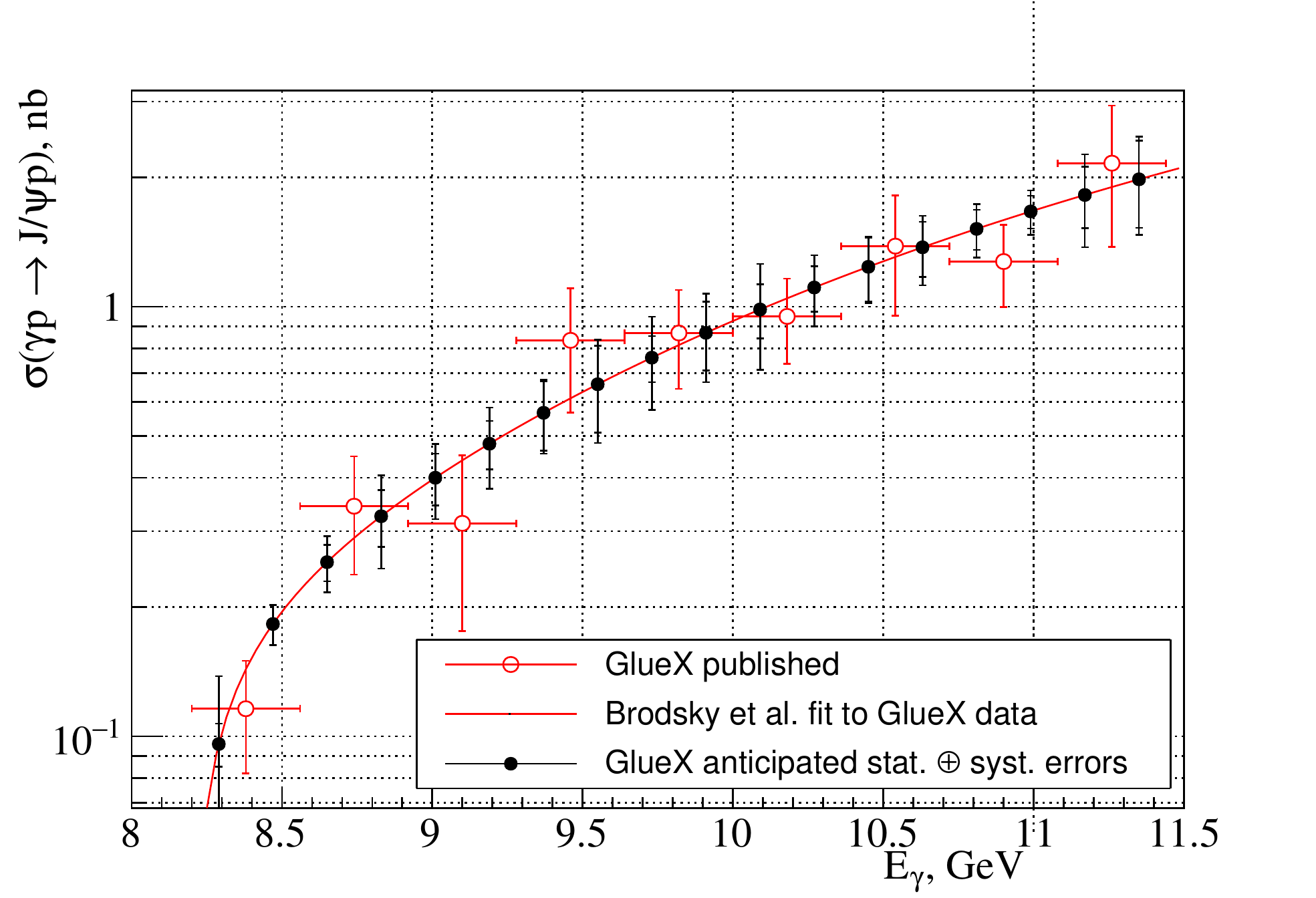}
        
        \caption{{\it Left}, Published GlueX results on the total cross section of $J/\psi$ photoproduction near threshold, along with various models of the cross section energy dependence~\cite{GlueX:2019mkq}.
        {\it Right}, Illustration of the expected uncertainties for the $J/\psi$ cross section results from the full GlueX-I data.  Measurements of the differential cross section $d\sigma / dt$ in several energy ranges are expected as well.
        The measurements in the left panel have an overall normalization uncertainty of 27\%.  This uncertainty is expected to be reduced for the measurements with the full GlueX-I data.}
        \label{fig:jpsiprl}
\end{figure*}

The maximum accessible photon energy at GlueX is 12 GeV, which allows for the photoproduction of all charmonium states up to the $\psi(3770)$ (and this last state only with a very small phase space).  
The first charmonium state identified at GlueX was the $J/\psi$, the study of which in photoproduction provides insight into the structure of the proton and a chance to search for the direct production of the $P_c$ states in the $J/\psi p$ channel~\cite{GlueX:2019mkq}.  The photoproduction cross section as a function of energy from roughly 25\% of the GlueX-I data shown in Fig.~\ref{fig:jpsiprl}~(left).  No evidence of $P_c$ production was seen, and model-dependent limits of $\mathcal{B}(P_c^+ \to J/\psi p)$ of less than a few percent were set. (say something about kinematic effects) However, these limits are consistent with models of the $P_c$ states as hadronic molecules, and other suggestions were made that would lead to looser limits, such as modifications of the production couplings or the dominance of charmed hadron loops~\cite{Du:2020bqj}.  Some of these hypotheses can be addressed with additional data, for example the model of Ref.~\cite{Du:2020bqj} predicts cusp-like structures in the the beam energy dependence of the total $J/\psi$ cross section, and updated results from the full GlueX-I run will soon be available.  These results will include 4 times as many $J/\psi$ as the published results, finer binning in beam energy, and the first measurement of the Mandalstam-$t$ dependence of this cross section in several beam energy regions, which will provide more sensitive studies of the production mechanisms and of $s$-channel resonance production. The projected uncertainties from these measurements are illustrated in Fig.~\ref{fig:jpsiprl}~(right). The full GlueX data promises to provide roughly a factor 3 more data for a total projected sample of around 6000~$J/\psi$.

The large acceptance and good particle identification capabilities of GlueX allow the identification of other charmonium states as well.  This will allow for the study of other facets of charmonium photoproduction, and the search for $P_c$ states that prefer to couple to channels beyond $J/\psi p$, and we summarize the prospects in the following (more details are given in Ref.~\cite{gluex_future}).  The outlook for identifying the spin-0 $\eta_c$ and $\chi_{c0}$ states is not good.  They preferentially decay into multi-pion states~\cite{pdg}, which are copiously produced in photoproduction.  For the $\eta_c$, the next most likely final states are $\eta\pi\pi$ and $K\overline{K}\pi$~\cite{pdg}, which happen to be important hybrid search channels, and no charmonium signals are seen.  Identification of these charmonium states must wait for an amplitude analysis or some other analysis development.  The $h_c(1P)$ decays about 50\% of the time to $\gamma \eta_c(1S)$~\cite{pdg}, where the $\approx500$~MeV transition photon gives some more hope of an identifiable signal, although one must compete with large backgrounds from hadronic decays including $\pi^0$'s.  The next most promising are the $\chi_{c1}(1P)$ and $\chi_{c2}(1P)$ states, which both have a strong decay to $\gamma J/\psi$.~\cite{pdg}  The $\psi(2S)$ also has a clean decay mode of $\pi^+\pi^- J/\psi$, although only a small amount of phase space remains for its production.  The photoproduction models described~\cite{Albaladejo:2020tzt} above predict very roughly that 50 $\chi_{c1}$ events and $<10$ $\psi(2S)$ events in these decay modes should be observable in the GlueX-I data.  Again, GlueX-II should provide an extra factor of 3 on top of this.  These statistics are small, but should allow for the first studies of the energy dependence of these cross sections, and the validation of these photoproduction models which also predict the production rates of $XYZP$ states~\cite{Albaladejo:2020tzt}.  

The study of open-charm final states is obviously a desirable next step, particularly to search for $P_c$ states that couple more strongly to the $D^{(*)}\Lambda_c^{(*)}$ and $D^{(*)}\Sigma_c^{(*)}$ channels.  However, GlueX relies heavily on exclusively reconstructed final states and kinematic fitting for improved mass resolution and background subtraction, and these charmed mesons decay in a large number of hadronic final states with branching fractions of a few percent each~\cite{pdg}.  Previously measured open charm cross sections are only an order of magnitude larger than that of $J/\psi$~\cite{Abe:1983pe}, and given the charmed hadron branching fractions an extra suppression factor of 25 for a given exclusive reaction is expected, it's clear that the full GlueX data is needed in order to possibly identify these final states.


\section{JLab Energy Upgrade: Charm Photoproduction Factory}

\begin{figure}
\begin{center}
\includegraphics[width=0.5\textwidth]{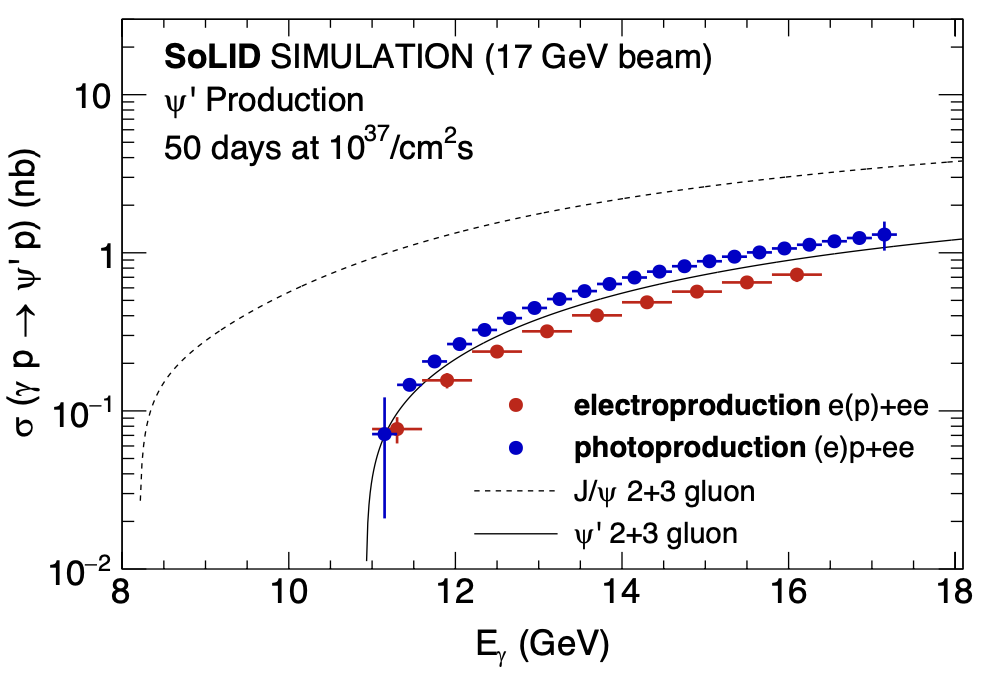}
\includegraphics[width= 0.37 \textwidth]{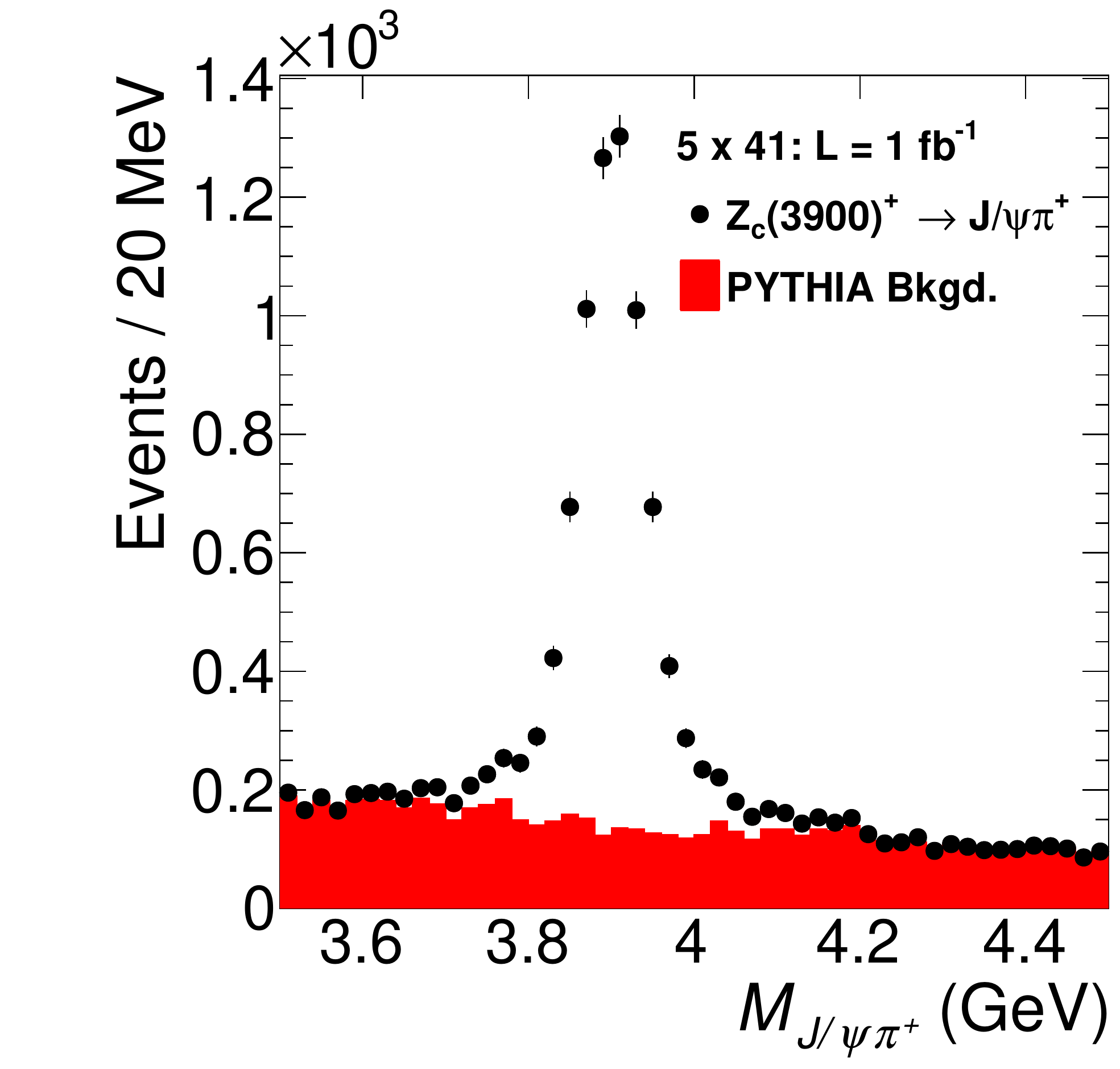}
\caption{\label{fig:upgrade} {\it Left}, Projections for SOLID $\psi(2S)$ total cross section vs beam energy for CEBAF upgrade with $E_e = 17$~GeV.  {\it Right}, Simulated $Z_c^+ \rightarrow J/\psi\pi^+$ mass distribution from smeared simulation for $5 \times 41$ GeV beam energies with contribution from PYTHIA background (red).}
\end{center}
\end{figure}

The existing CEBAF accelerator at Jefferson Lab provides an electron beam with energies up to 12 GeV and fixed target luminosities up to $10^{39}$~cm$^{-2}$s$^{-1}$. A proposed energy, described in Ref.~\cite{Arrington:2021alx} would extend the accessible energy to 20-24~GeV using Fixed Field Alternating Gradient in the existing recirculation arcs with similar luminosity to the existing CEBAF accelerator.  Such a facility with up to 24 GeV in photon beam energy will provide new opportunities to study conventional charmonium by significantly extend the energy range of the $J/\psi$ photoproduction cross section measurements described in Sec.~\ref{sec:gluex}.  The additional energy also allows for the first measurements of threshold $\psi(2S)$ production, to test the universality of threshold charmonium production.  Simulations from the SoLID detector proposal, utilizing a 17 GeV $e-$ beam show excellent precision in determining the photo- and electro-production cross sections of the $\psi(2S)$, as shown in Fig.~\ref{fig:upgrade} (left).

Looking beyond conventional charmonium this upgrade would provide the energy needed to search for exotic states in the exclusive photoproduction reactions described in Sec.~\ref{sec:jpac}.  While the diffractive production of vector charmonia, such as the $J/\psi$ and $\psi(2S)$, increases rapidly with energy above threshold, the production of other possible exotic states, such as $X$ and $Z$ have a significant enhancement near threshold as shown in Fig.~\ref{fig:Zplots}.  Therefore, a high-luminosity, fixed-target experiment with a 24 GeV photon beam is an ideal place to study the threshold production for charmonium states with masses up to 5.5 GeV.  Such measurements would be complementary to the observations so far at hadron and $e^+e^-$ colliders by eliminating some of the possible kinematical effects in production, as well as providing new observables of through beam and target polarization to understand the various production mechanisms.

\section{EIC: High energy charm and bottom quark photoproduction}

The Electron Ion Collider (EIC) is a high-luminosity polarized $ep$ and $eA$ collider with variable center-of-mass energies from 20-140 GeV.  A broad physics program is outlined in the EIC white paper~\cite{Accardi:2012qut} including key questions in particle and nuclear physics on the 3-dimensional structure of the nucleon and high parton densities in nuclei.  Recently a hadron spectroscopy program at the EIC has developed through theoretical predictions for $XYZ$ photoproduction cross sections and simulations of experimental requirements, described in detail in the EIC Yellow Report~\cite{AbdulKhalek:2021gbh} and summarized here.  While the variable center-of-mass energy of the EIC provides access to the threshold production of $X$ and $Z$ states, the higher energy is ideal to study the diffractively-produced $Y$ states whose cross sections are expected to rise significantly with energy where the highest instantaneous luminosities are achieved.

The spectroscopy of unconventional quarkonia in photoproduction relies on the efficient detection of all the meson decay products and adequate resolutions to identify and study the produced resonances.  An example study of a particular reaction of interest $\gamma p \rightarrow Z_c(3900)^+ n$ provides useful insight to the detector requirements in $ep$ collisions.  Here we will focus on the $Z_c(3900)^+ \rightarrow J/\psi\pi^+$, with $J/\psi \rightarrow e^+e^-$ .  This reaction is simulated using the amplitudes described in Sec.~\ref{sec:jpac} which were integrated into the \textsc{elSpectro} event generator ~\cite{GitHub:elspectro}.

The asymmetric collider kinematics of the EIC requires a rather specialized detector system to identify both the decay products of the produced meson system, but also tag the scattered beam particles $e'$ and $p'$ to ensure the exclusive reconstruction of the full final state.  The $XYZ$ system's decay products often receive a considerable boost in the proton beam direction, requiring detector acceptance up to a pseudorapity of 3.5.  With considerable focus on hadronic final states for many EIC measurements identification of the $J/\psi \rightarrow e^+e^-$ decay imposes additional particle identification requirements on the detector systems. The $J/\psi\pi^+$ mass distribution in Fig.~\ref{fig:upgrade} illustrates the level of background (red) and expected $Z_c(3900)^+$ signal (black) assuming a nominal electron-pion separation with inclusive hadron production was simulated using PYTHIA.  

With expected integrated luminosities expected for the EIC on the order of 1-10~fb$^{-1}$ there are new opportunities to study rare exclusive processes not accessible at HERA.  In addition, measurements of conventional quarkonium states have shown that nuclear breakup effects are dependent on the radius of the observed state~\cite{Leitch:1999ea, Alessandro:2006jt, Arleo:1999af}.  It is anticipated that similar suppression effects for exotic hadrons can be used to discriminate between compact multiquark and molecular models of their structure.  This has been studied in high multiplicity collisions~\cite{Cho:2010db, Cho:2017dcy, Zhang:2020dwn, Wu:2020zbx, Esposito:2020ywk,Braaten:2020iqw} and could continue in $eA$ collisions with varying target nuclei at the EIC.

\section*{Acknowledgements}

This work was supported by the U.S.~Department of Energy under contract DE-AC05-06OR23177 under which Jefferson Science Associates, LLC, manages and operates Jefferson Lab, DE-FG02-87ER40365 at Indiana University, DE-FG02-92ER40735 at Florida State University, and Early Career Award contract DE-SC0018224. 
We also acknowledge support from 
Spanish Ministerio de Econom\'ia y Competitividad and  
Ministerio de Ciencia e Innovaci\'on under Grants 
No.~PID2019-105439G-C22,
and No.~PID2020-112777GB-I00 (Ref.~10.13039/501100011033),
Polish Science Center (NCN) under Grant No.~2018/29/B/ST2/02576,
UNAM-PAPIIT under Grant No.~IN106921,
and CONACYT under Grant No.~A1-S-21389. 
%
MA is supported by Generalitat Valenciana under Grant No.~CIDEGENT/2020/002.
CFR is supported by Spanish Ministerio de Educaci\'on y Formaci\'on Profesional under Grant No.~BG20/00133.
VM is a Serra Húnter fellow and acknowledges support from the Spanish national Grant No. PID2019–106080 GB-C21 and PID2020-118758GB-I00.
ANHB is supported by the Deutsche Forschungsgemeinschaft (DFG) through the Research Unit FOR
2926 (project number 40824754).







\end{document}